\documentclass[11pt,a4paper]{article}
  \usepackage{jheppub}
\pdfoutput=1

\usepackage{fancyhdr}
\usepackage{amsthm}
\usepackage{enumerate}
\theoremstyle{definition}
\usepackage{graphicx,rotating,booktabs}
\usepackage{subfig}
\usepackage{tikz}
\usetikzlibrary{shapes,backgrounds}

\newcommand{\MET}{{\not\!\! E}_T}

\usetikzlibrary{decorations.pathmorphing}
\usetikzlibrary{decorations.markings}
 \tikzset{
    photon/.style={decorate, decoration={snake}, draw=red},
    particle/.style={draw=blue, postaction={decorate},
        decoration={markings,mark=at position .5 with {\arrow[draw=blue]{>}}}},
    antiparticle/.style={draw=black,dashed},
    gluon/.style={decorate, draw=black,
        decoration={coil,amplitude=4pt, segment length=5pt}}
     }
\usepackage{slashed}

\def\beq{\begin{equation}}
\def\eeq{\end{equation}}
\def\bea{\begin{eqnarray}}
\def\eea{\end{eqnarray}}
\def\beu{\begin{quotation}}
\def\eeu{\end{quotation}}

\title{Long-Lived Particle Searches in R-Parity Violating MSSM}

\author[a,b]{ Nosiphiwo Zwane}
\affiliation[a]{Perimeter Institute for Theoretical Physics, 31 Caroline St. N., Waterloo, ON, N2L 2Y5, Canada}
\affiliation[b]{Department of Physics and Astronomy, University of Waterloo, Waterloo, ON, N2L 3G1, Canada}

\abstract{In this paper we study the constraints on MSSM  R-Parity violating decays when the Lightest Superpartner (LSP) is moderately long lived. In this scenario the LSP vertex displacement may be observed at the LHC.  We compute limits on the RPV Yukawa couplings for which the vertex displacement signature maybe used. We then use ATLAS and CMS displaced vertex , meta-stable and prompt decay searches  to rule out a region of sparticle masses.   
}

\begin{document}

\date{\today}
\maketitle
\flushbottom

\section{Introduction}

The discovery of a Higgs boson at the LHC \cite{ATLASHiggs,CMSHiggs} sharpens the long-standing puzzle of the electroweak hierarchy.  Generic new physics above the weak scale would induce large radiative corrections to the Higgs mass, so that a precise fine-tuning of classical and quantum contributions to the Higgs mass must be invoked to explain the relative smallness of the weak scale.  To avoid such a tuning, the Universe must have some means of protecting the weak scale --- either \emph{all} physical phenomena (including an ultraviolet completion of gravity, dark matter, and the generation of neutrino masses and a baryon asymmetry) occur at or below the weak scale, as in extra-diemensional models with a low fundamental scale \cite{ADD}, the Higgs boson itself is composite \cite{Kaplan:1983sm,Agashe:2004rs}, or a fundamental symmetry that protects the Higgs mass from radiative corrections is restored near the weak scale.  Concrete realizations of all of these proposals predict new particles near the weak scale  and/or modified Higgs properties, which to date have not been observed.  Nonetheless, the dramatic fine-tuning implied by the alternative motivates a continued search for new TeV-scale physics, which must be broad enough to test all incarnations of known models to address the hierarchy problem.

Of the many symmetries proposed to stabilize the electroweak hierarchy, supersymmetry stands out as particularly attractive: it allows a perturbative completion of the Standard Model (SM) up to the Planck scale, and TeV-scale superpartners modify the running of SM gauge couplings so that they unify at a scale $M_G \sim 10^{16} \mbox{ GeV}$.  

Unlike the SM, supersymmetric completions do not have baryon number ($B$) and lepton number ($L$) as accidental symmetries --- indeed, the Minimal Supersymmetric Standard Model (MSSM) particle content admits two types of marginal interactions that violate $L$ and one that violates $B$\footnote{A further $L$-violating relevant operator, the superpotential coupling $\mu' L H_d$, can be removed by a field redefinition and we do not discuss it further.}.  Since the combination of unsuppressed $B$- and $L$-violation would lead to rapid proton decay, some or all of these terms must be very small --- either by a new symmetry like R-parity that forbids all three terms, by a perturbative $B$ or $L$ symmetry, or by a flavor structure that suppresses first-generation couplings sufficiently to evade present bounds on proton decay \cite{MFV}.  

The presence or absence of R-parity-violating (RPV) interactions has dramatic conseqences for the LHC signals of supersymmetry.  Exact R-parity symmetry guarantees that the lightest superpartner (LSP) is exactly stable; any superpartners produced at a collider will therefore decay into an LSP, which in turn escapes the detector without interacting.  The resulting missing transverse energy ($\MET$) signature is one of the most used discriminating variables in collider searches for supersymmetry.  RPV interactions lead instead to the decay of the LSP, which can occur within the detector volume for couplings as small as $10^{-9}$ depending on the identity of the LSP.  If $L$-violation dominates (for example, if a perturbative $B$ symmetry is assumed to guarantee the proton's stability), these decays produce additional leptons.  If instead $B$-violation dominates (for example, if a perturbative $L$ symmetry or minimal flavor-violating structure guarantee the proton's stability), the $\MET$ signature from stable LSPs is replaced by vertex signature with additional jets in each SUSY pair-production event.  These are typically much harder to detect than missing energy or lepton-rich decays \cite{Lisanti:2011tm,Evans:2013jna}, but are increasingly targeted by dedicated LHC searches (see e.g. \cite{CMS:2014exa,Aad:2015lea}) so that constraints on colored superpartners are still at the level of $700-1000$ GeV for most scenarios. 

This work investigates the sensitivity of present LHC searches to $\sim TeV$-mass superpartner production in a variety of simplified models, when the LSP decays via $B$-violating RPV interactions.  Similar scenarios have been studied in \cite{Brust:2012uf,Graham:2014vya,Cui:2014twa}.
Our main focus here is on studying the effects of LSP decays with mm to meter-scale displaced vertices, which can arise for RPV couplings between $\sim 10^{-3}$ and $\sim 10^{-9}$, depending on the nature of the LSP.  
Our reasons for studying this displaced parameter region are threefold: 
\begin{itemize}
\item Displaced decays occur in a significant fraction of the B-violating coupling parameter-space, where the collider signals are quite different from either prompt decays or collider-stable LSPs.
\item Di-nucleon decay constrains the $B$-violating coupling $\lambda''_{121} \lesssim 5 \times 10^{-7} \left( \frac{m_{\tilde{g}}}{1\text{TeV}}\right)^{1/2} \left( \frac{m_{\tilde{q}}}{1\text{TeV}}\right)^{2}  $.  If $B$-violating couplings have an anarchic flavor structure, these constraints strongly favor the parameter regions where either RPV decays are displaced or the LSP is collider-stable.  
\item The possible connection between $B$-violation in RPV decays and the baryon asymmetry of the Universe is an interesting one.  On one hand, late-decaying LSPs decay out of equilibrium, and so may be relevant to the third Sakharov condition \cite{Cui:2012jh,Cui:2013bta}. On the other hand, if baryogenesis occurs above the weak scale then long-lived LSPs are typically \emph{required} to avoid washing out the asymmetry at the weak scale \cite{Barry:2013nva}. 
\end{itemize}

LHC searches for exotic late-decaying particles typically involve modified object reconstruction algorithms and cuts on ``low-level'' variables, which are nearly impossible to model outside a full detector simulation.  We have not attempted this modeling, which therefore introduces a significant source of uncertainty in our analysis.  Instead, we have roughly estimated signal rates either by rescaling published limits according to a lifetime-dependent probably of decaying in a search's region of interest, or by a naive combination of NLO production cross-sections, lifetime-dependent decay probabilities, and parametrized efficiencies where possible.  For consistency, we adopt the same procedure even for the prompt and collider-stable analyses where a more detailed simulation in a ``generic'' detector simulation like PGS \cite{PGS} or Delphes \cite{deFavereau:2013fsa} would be possible.  This procedure considerably limits the reliability of our model-exclusion contours --- they are certainly less accurate than the standard of modern simulation would allow, and in some cases may be inaccurate by up to an order of magnitude in rate.  Even so, since the parameter space of interest spans many decades in lifetime and pair-production cross-sections fall as a high power of superpartner mass, there are qualitative lessons to be learned from such a study.

In particular, as has been found by other studies, the constraints on strongly-interacting superpartners with prompt RPV decays exclude gluino masses up to $\sim 700-1000$ GeV --- they are weaker than when the LSP (colored or neutral) is stable, but already strong enough to rule out a truly natural superpartner spectrum.  
Unsurprisingly, displaced vertex searches --- which have negligible physics backgrounds and small instrumental backgrounds --- have excellent comparative sensitivity.  These dedicated DV searches therefore have impressive discovery potential for supersymmetric theories that give rise to displaced decays.  They are also \emph{essential} to fully testing the hypothesis of TeV-scale supersymmetry, because significantly displaced decays are often rejected by the event quality requirements in non-DV searches (see e.g. \cite{ATLAS-CONF-2014-037}).  An LSP with proper lifetime in the millimeter-to-meter range is \emph{only} efficiently probed by DV searches.  The attention given to such searches, while increasing in recent years, remains far less than for the complementary missing energy searches. 

Towards completion of this work, closely related analysis \cite{sp1,sp2} appeared which overlap with this work. They both do analysis for stops, gluino and Higgs LSP.  In our work we include analysis for a Bino LSP from gluino also a Bino LSP from squark ,as well as stops and gluino LSP.

The paper is outlined as follows: In section 2 we review the physics of R-parity violation and existing low-energy constraints \cite{minireview,MFV,myrpv}. In section 3, we summarize the LHC searches used in our analysis, then estimate the constraints on simplified models with stop, gluino, and bino LSPs as a function of RPV coupling and masses, for either hierarchical (3rd-generation-dominated) or universal RPV couplings.  

\section{R-Parity Violation}
In this section, we briefly review the structure of the MSSM with and without conserved R-parity and low-energy constraints on R-parity violation, following \cite{SUSY,myrpv}.
The most general gauge-invariant and renormalizable superpotential for the particle content of the  MSSM is
\begin{align}
W_{MSSM}&= \tilde{\bar{u}}\textbf{Y}_u\tilde{Q}H_u - \tilde{\bar{d}}\textbf{Y}_d\tilde{Q}H_d - \tilde{\bar{e}}\textbf{Y}_e\tilde{L}H_d + \mu H_uH_d  \label{mg} \\
 &\, + \frac{1}{2}\lambda ^{ijk}L_iL_j\bar{e}_k + \lambda ^{'ijk} L_iQ_j\bar{d}_k + \mu ^{'i}L_iH_u + \frac{1}{2} \lambda ^{''ijk}\bar{u}_i\bar{d}\bar{d}_k \label{RPVP}
\end{align}
The last four terms violate either baryon number $B$ or lepton number $L$.  It is conventional to remove the $\mu^{'i}$ terms by a field redefinition. The other three $B$- or $L$-violating terms can be forbidden by assuming that ``R-parity'', a discrete $\mathbb{Z}_2$ symmetry with parity assignments
\begin{equation} \nonumber
P_{R} = (-1)^{3(B-L)+2S}
\end{equation}
where $S$ is the spin of the particle. Thus all SM particles have $P_{R}=1$ and all the SM superpartners (sparticles) have $P_{R}=-1$. If R-parity is conserved then each vertex involves an even number of sparticles, and hence:
\begin{enumerate}
\item The lightest superpartner (LSP) is stable since it can not decay to light quarks as this would lead to an odd number of sparticles at that vertex. The LSP if its neutral makes a good candidate for dark matter (DM).
\item Each sparticle other than the LSP should decay to an odd number of LSPs (usually just one) plus additional SM particles.  These decays can proceed entirely through interactions related by supersymmetry to SM interactions, which have ${\cal O}(1)$ dimensionless couplings and are prompt for generic superpartner mass splittings.
\item High-energy collider reactions always produce an even number of sparticles.
\end{enumerate}

Searches at the LHC and past colliders have not found superpartners, but continue to set lower limits on their masses in various scenarios. 
For example, ``jets + $\MET$'' searches \cite{CMSOut} currently exclude gluinos and squarks (of the first two generations) below
1.2 TeV and 780 GeV respectively in simplified models containing only these squarks, a gluino octet and a relatively light stable neutralino. 
In MSUGRA models with $\tan \beta $ = 30, $A_0 = -2m_0$  and $\mu >$ 0, squarks and gluinos of equal mass are excluded for masses below 1700 GeV \cite{ATLASJ2}. The lightest stop, which contributes directly to the fine-tuning of the Higgs mass, is constrained to be heavier than about 750 GeV for much lighter LSPs, with the exception of small windows with stop-LSP mass splittings near $m_t$ or $m_W$.  

These constraints rely crucially, however, on the stability of the neutralino LSP.  Their growing severity motivates further exploration of supersymmetric models where R-parity is not an exact symmetry.  If the R-parity-violating (RPV) couplings are small, then properties 2.~and 3.~above remain true to a good approximation, but 1.~is severely altered.  Rather than being exactly stable, the LSP can decay to SM particles through the couplings $\lambda$, $\lambda'$ and $\lambda''$ in \eqref{RPVP}.  Thus the LSP is not a dark matter candidate in RPV models, and the phenomenology of superpartner production at colliders is dramatically altered by the LSP decay.

\subsection{Low-Energy Constraints on RPV SUSY}
\label{sec}
As we have already noted, the R-parity violating couplings also break the baryon and lepton number symmetries of the Standard Model.  Tests of these symmetries therefore set powerful constraints on the RPV couplings, discussed in \cite{myrpv,KaoTakeuchi}.  

\begin{enumerate}
\item \textbf{Proton and Neutron Decays:} Proton and neutron decay rates have been bounded at the level of of $10^{-33}$ to $10^{-34}/\rm{year}$ for a wide variety of decay modes.  These bounds imply incredibly severe constraints on the products of $B$- and $L$-violating RPV couplings, at the level of $\lambda' \lambda'' \lesssim 10^{-24}$ 
 for few-TeV superpartners  \cite{minireview} 
 Several scenarios discussed in the literature naturally evade these constraints by violating either lepton or baryon number, but not both (or violating one to an exponentially small degree as in \cite{MFV}). Our focus here is on scenarios with $B$-violating RPV interactions (i.e. $\lambda''\neq 0$), so we assume from here on that this constraint is satisfied by virtue of the $L$-violating couplings $\lambda$ and $\lambda'$ being vanishingly small.
We further assume a spectrum with no gravitino or axino lighter than the proton, as these would allow proton decays through $\lambda''$ couplings only.


\item \textbf{Dinucleon decay:} The strongest constraint involving only $\lambda ''$ couplings comes from di-nucleon decay. 
  Sher \textit{et. al} \cite{SherK} set bounds on $\lambda_{121}''$ from the partial width for the direct transition $NN \rightarrow K^+ K^-$. The decay rate for this process is 
\begin{align}
\Gamma \sim \rho_N \frac{128 \pi \alpha_s^2 \lambda_{121}^4 \Lambda^{10}}{M_N^2 M_{\tilde{g}}^2 M_{\tilde{q}}^8}
\end{align} 
Using the lifetime bound from Litos \textit{et. al}\cite{newNN}, $\tau \geqslant 1.7 \times 10^{32}$, $\rho_N = 0.25 fm^{-3},$ $\alpha_s \sim 0.12 $ and $ \Lambda \sim 500 - 700$ MeV, giving the bound,

\begin{align*}
\lambda''_{121}  \lesssim 5 \times 10^{-7} \left( \frac{m_{\tilde{g}}}{1\text{TeV}}\right)^{1/2} \left( \frac{m_{\tilde{q}}}{1\text{TeV}}\right)^{2} .
\end{align*}

\item \textbf{UV perturbativity:}  Barbier \textit{et. al} \cite{myrpv} set bounds on the RPV couplings using the RGE by imposing that the ultraviolet scale evolution be perturbative up to the large unification scale.

\item \textbf{$n-\bar{n}$ oscillation:}  Two types of diagram contribute to $n-\bar{n}$ oscillation:  tree diagrams that involve unknown squark mass-mixing \cite{SUSY,Chem}  and  box diagrams independent of squark mixing \cite{SherK}.  Remaining agnostic about the squark-mass flavour structure, will do not impose the former constraint.   The latter sets only a weak bound which we will not consider in this paper.

\end{enumerate}

\subsection{Patterns of R-Parity Violation}
The are many possibilities as to how R-parity can be broken. For example, the minimal flavor violation (MFV) model \cite{MFV} assumes an MFV symmetry in the MSSM instead of assuming mass universality and R-parity. The RPV couplings must be proportional to appropriate products of Yukawa couplings respecting this flavor symmetry, for example, the B-violating coupling \cite{MFV},
\begin{align}
\lambda ''_{ijk} &= w''y_i^{(u)}y_j^{(d)}y_k^{(d)}\epsilon _{jkl}V_{il}^* ,  
\end{align}
where the $y$'s are Yukawa couplings, $w''$ an arbitrary coefficient and $V_{il}$ is the CKM matrix. The largest coupling by far is $\lambda _{tsb}''$. Models of this kind we shall call hierarchical RPV models. 

Other models have all couplings $\lambda_{ijk}''$ of the same order. As a benchmark we consider the case of universal couplings $\lambda_{ijk}'' = \lambda ''$ for all $ijk$. These scenarios are in general more constrained than the hierarchical models. They also have different phenomenology at the LHC, particularly for gaugino LSPs.

\section{Lightest Superpartner LSP}

In RPV SUSY, a single sparticle may be produced. Since we take the RPV Yukawa coupling to be very small, these will be rare and can be neglected. 
In this paper, production of the superpartners is considered to be R-parity conserving so that there will be no single sparticle production. Superpartners that are not the LSP  are also taken to decay via R-parity preserving channels. Only the LSP decays via RPV couplings.


The LSP can be long lived if the RPV Yukawa couplings are very small. In this case, each LSP decay products originate from a point in space that can be resolved from the primary vertex. This vertex displacement may be observed in a collider detector. A detector has a minimum displacement and a maximum displacement that can be observed. When the lifetime is too small the displacement may be too small hence not observable. When the lifetime is too large the LSP may decay outside the detector resulting to missing energy signature.

The values of the minimum and maximum observable displacement differs for different detectors and analysis. These depend on the geometry and performance of a specific detector, and on which criteria is used to reconstruct vertices and to select events.  

This paper considers three possible LSPs: squark, gluino and bino. The lifetime of the LSP is calculated to set limits on the RPV couplings $\lambda ''$ that lead to observable displacement. Results from ATLAS and CMS searches in Table \ref{table2}  are used to exclude regions where RPV decays can not occur.  

\subsection{Relevant LHC Searches}
To compare with experiment the number of expected events for a pair of LSPs decaying inside the detector  is used. If the number of expected events is greater than the maximal signal allowed by an analysis, that decay is ruled out. The number of events for a pair of LSPs decaying with an observable vertex displacement is

\begin{align}
N &=  \mathcal{L} \times \sigma \times P_{\text{decay}}^2 \times P_{\text{vertex}}^2 \times P_{\text{trigger}},  \label{Nevents} 
\end{align}
where $\sigma$ is the cross section production of the pair of sparticle from a proton-proton interaction, $\mathcal{L}$ is the integrated luminosity, $P_{\text{vertex}}$ is the efficiency of reconstructing the vertex and $P_{\text{trigger}}$ is efficiency of the trigger. $P_{\text{decay}}$ is the probability that the LSP decays with an observable vertex displacement and is given by

\begin{align}
P_{\text{decay}} = \int_{l_{min}}^{l_{max}} \frac{1}{\gamma \beta \tau c} \quad e^{-\frac{l}{\gamma \beta c \tau}} dl,
\end{align}
where $l_{min}$ and $l_{max}$ are the minimum and maximum vertex displacement that can be observed in the collider, $\gamma \beta$ is the Lorentz boost, $c$ the speed of light, $\tau$ is the lifetime of the LSP. To normalize the probability we assume that given that a LSP is produced and the RPV couplings are non-zero then the probability that it decays with a vertex displacement between 0 and $\infty$ is 1. If only one of the LSP decays within the detector then the number of events is given by
\begin{align}
N &=  2 \mathcal{L} \times \sigma \times P_{\text{decay}}P_{\text{vertex}} (1- P_{\text{decay}} P_{\text{vertex}})  P_{\text{trigger}} \label{Nevents2}.
\end{align}


\begin{sidewaystable}[h!]
\caption{Relevant LHC Searches}
\begin{small}
\begin{tabular}{|c|l|c|c|c|c|}
\hline &Search & $\mathcal{L}$ & $l_{min}$  & $l_{max}$& $\sigma $  \\[6pt] 
\hline (a) Displaced Vertex Searches &  Vd1 CMS 2013 \cite{CMSvtx} & 18.6 $fb^{-1}$ & 0.1 cm & 20 cm & 0.3 fb - 300 fb \\[6pt]
\cline{2-6} &  Vd2 ATLAS 2012 \cite{ATLAS2012}  & 1.94 $fb^{-1}$ & 4.5 m & 7 m & 16 pb - 24 fb \\ [6pt]
\cline{2-6} &  Vd3 ATLAS using muon trigger \cite{ATLASmu2}  & 20.3 $fb^{-1}$ & 4mm & 180mm & $0.15912$ pb \\ [6pt]
\hline (b) Meta-Stable LSP Searches &   Od1 CMS search for R-hadrons \cite{CMSR2}  & 18.8 $fb^{-1}$ & 1.2 m & $\infty$ &  20 pb - $ 4\times 10^{-4}$pb (200 GeV - 1000 GeV)\\ 
\cline{2-6} &  Od2 ATLAS search for jets   & 20.3 $fb^{-1}$ & 11m & $\infty$  &  $16$ fb   \\
& {and missing energy \cite{ATLASJ2} } & & & & \\ 
\cline{2-6} &  Od3 ATLAS search for jets   & 19.5 $fb^{-1}$ & 7.5m & $\infty$  &  $1-10^{-3}$ pb   \\ 
& {and missing energy \cite{CMSOut} } & & & & \\ 
\cline{2-6} &  Od4 ATLAS search for R-hadrons \cite{ATLASR}  & 4.7 $fb^{-1}$  & 11m & $\infty$ & 0.7 pb - $1.5 \times 10^{-3}$pb (200 GeV - 1000 GeV)\\ [6pt]
\hline Prompt Decay Searches &  J1 CMS dijets \cite{CMSs} & 12.4 $fb^{-1}$ & 0m & 1mm & 0.2 pb - $5 \times 10^{-3}$pb (300 GeV - 1200 GeV)\\ [6pt]
\cline{2-6}  & J2 ATLAS search 3 jets \cite{ATLAS3jetb,ATLAS3jetb2,CMSb,CMS3jetb} & 20.3 $fb^{-1}$& 0m & 1mm & 18700 pb - 0.0572 pb (100 GeV - 800 GeV)\\ [6pt]
\cline{2-6} &  SSL (same sign lepton) ATLAS search \cite{Asame} & 21 $fb^{-1}$ & 0 & 0.4mm & 0.34 fb \\
\hline  Limits from section \ref{sec} & IN1 Dinucleon decay  &\\
\cline{2-6}  & IN2 RGE   & \\
\hline
\end{tabular} \label{table2}
\end{small}
\end{sidewaystable}

\begin{itemize}
\item[(a)]{Displaced Vertex Searches }
\begin{itemize}
\item[Vd1] The CMS detector \cite{CMSvtx,newCMS01} was used to search for a heavy scalar particle which decays to a pair of long-lived massive neutral particles decaying to quark-antiquark pair. They searched for a pair of jets originating from a displaced vertex. Only one of the long-lived massive neutral particles must have a secondary vertex reconstructed, the other can decay anywhere. The trigger efficiency is taken to be 40\%.

\item[Vd2] A vertex displacement search was also done using the ATLAS detector \cite{ATLAS2012}. The analysis were based on a search for a light Higgs decaying into two weakly interacting long-lived pseudo-scalars from a proton-proton collision. They looked for two isolated back to back vertexes in the muon spectrometer and found no signal. Both of the long-lived particles are expected to decay within the muon spectrometer to give such a signal. Trigger efficiency is taken to be 40\% and the efficiency in reconstructing the vertex to be 40\%.

\item[Vd3] A search for long-lived particle using displaced vertices was done using the ATLAS detector \cite{ATLASmu,ATLASmu2} and no signal was found. The search used a muon trigger, and assumes that the long-lived particle decays to a muon. They searched for a muon and many charged tracks originating from a single vertex. In the model we consider, this can arise from LSP decaying into top quarks, which produces an isolated muon with $10\%$ probability. 
\end{itemize}
\item[(b)]{Meta-Stable LSP Searches (LSP Decaying Outside Detector)}
\begin{itemize}
\item[Od1] A search for coloured long-lived particles ($\tilde{g}$ or $\tilde{q}$) that hadronise forming R-hadrons was performed using the ATLAS detector at LHC \cite{ATLASR,ATLASR3}. Sparticles were assumed to be stable within the ATLAS detector and be observed as heavy stable charged particles. Data matched Standard Model background, hence a long-lived gluino LSP (stop LSP) was excluded up to a mass of 985 GeV (683 GeV). CMS also did a R-hadron search \cite{CMSR,CMSR2} which excluding long-lived gluino (stop) LSP masses of 1322 GeV (820 GeV). These searches also apply, with reduced efficiency, to shorter LSP lifetimes such that the LSP leaves the detector in a fraction of events. 

\item[Od2] ATLAS \cite{ATLASBino,ATLASJ2} searched for squarks and gluinos in final states containing jets, missing transverse momentum with no high-pT electrons nor muons. These searches are relevant to long-lived neutralino LSP. The E-cuts (which require 6-jets and $E^{\text{miss}}_T > 160$GeV) and loose selection in table 4 of \cite{ATLASJ2} are used in this paper with the limits in Figure 19 of \cite{ATLASJ2}. 

\item[Od3] ATLAS \cite{CMSOut} searched for squarks and gluinos with multijets in final states and large missing transverse momentum. This search applies to a pair of gluinos (or squarks) decaying to a bino LSP that does not decay inside the detector. Six jets with large missing energy are expected in this case. Limit from figure 7 \cite{CMSOut} are used. 
\end{itemize}

\item[(c)]{Prompt Decay Searches }
\begin{itemize}
\item[J1] The CMS detector \cite{CMSs,newCMS02} was used to search for new colored particles, produced strongly in pairs, that decay hadronically to dijets from a proton-proton collision. This search did not use the vertex displacement signature but used resonance signature looking for four high transverse momentum $p_{T}$  jets and found no signal. This search excluded a stop with mass between $200$ GeV and $385$ GeV.
 
\item[J2] ATLAS searched for a pair of 3 jets in pp collision \cite{ATLAS3jetb}, using an RPV gluino-decay signal as a benchmark model. They used two analyses; the resolved analysis and the boosted analysis. In this paper we consider results from resolved analysis because they are more constrained. For resolved analysis gluino masses are excluded up to $639$GeV. CMS also performed a 3-jets search \cite{CMS3jetb}, which excluded gluino masses below 460 GeV. ATLAS \cite{ATLAS3jetb2} also performed another search for massive particle decaying to multiple jets. In this search a pair-produced gluinos each decay into three light-flavor quarks, ruling out $m_{\tilde{g}} < 971$ GeV.


\item[SSL] ATLAS searched for same sign leptons \cite{Asame,CMSsame2}, and no excess signal was observed. The search was used to set limits on a 
RPV model, $\tilde{g} \rightarrow t \,\tilde{t} \rightarrow t\,s\, b$ , with the coupling $\lambda_{323}=1$. This excluded gluino masses below about 860 GeV over the whole range of considered top squark masses (up to 1 TeV). We used SR3b analysis, where events are required to have atleast 3 b-jets.
\end{itemize}

\end{itemize}

Many features of the experimental analyses, particularly for the displaced vertex searches, are difficult to simulate outside the full analysis framework. This introduces uncertainties as large as a factor of 2-4 in rates. However the parameter space of interest spans so many orders of magnitude in proper life time and a few orders of magnitude in cross sections so that even with these uncertainties we can still draw relevant conclusions. Recent vertex displacement searches \cite{late1,late2} have not been included in this paper.


\subsection{Squark LSP}
\begin{figure}[h!]
\centering
\begin{tikzpicture}
  \coordinate (a) at (2,2) {};
  \coordinate (b) at (6,2) {};
  \coordinate (c) at (10,0) {};
  \coordinate (d) at (10,3) {};
  \draw [antiparticle] (a) -- (b)node[midway,above] {$\tilde{q}_i$};
  \draw [particle] (b) -- (c)node[midway,above] {$q_j$};
  \draw [particle] (b) -- (d)node[midway,right] {$q_k$};
\end{tikzpicture} \label{squark}
\caption{Squark LSP}
\end{figure}
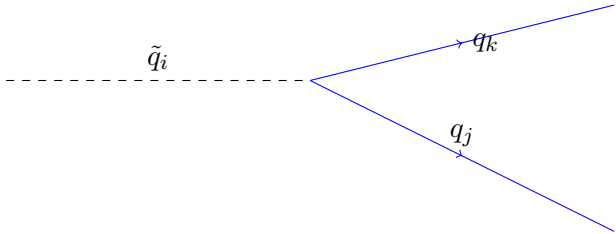

We first consider the case when the LSP is a stop. It decays through the RPV interaction,
\begin{align}
\mathcal{L} &\supset \lambda ^{''}_{3jk} \tilde{\bar{t}}_{R}\bar{d}_{j,R}\bar{d}_{k,R}.
\end{align}
The stop is expected to be produced in pairs since it will be produced via a R-parity conserving decay thus we expect this decay to have four jets no missing energy nor isolated leptons. We shall consider two cases: universal RPV coupling $\lambda ''$ and for hierarchical RPV couplings $\lambda ''_{323}$ because it is the largest coupling, hence has the largest branching ratio. The stop lifetime is 

\begin{equation}
\tau = \frac{16\pi}{\lambda ^{''2} _{ijk} m_{\tilde{t}}},
\end{equation}
for hierarchical RPV coupling and 
\begin{equation}
\tau = \frac{16\pi}{3\lambda ^{''2} m_{\tilde{t}}},
\end{equation}
for universal RPV coupling. \\

\begin{figure}[h!]
\centering
\subfloat[ Assuming universal couplings.]{
\includegraphics[width=0.45\textwidth]{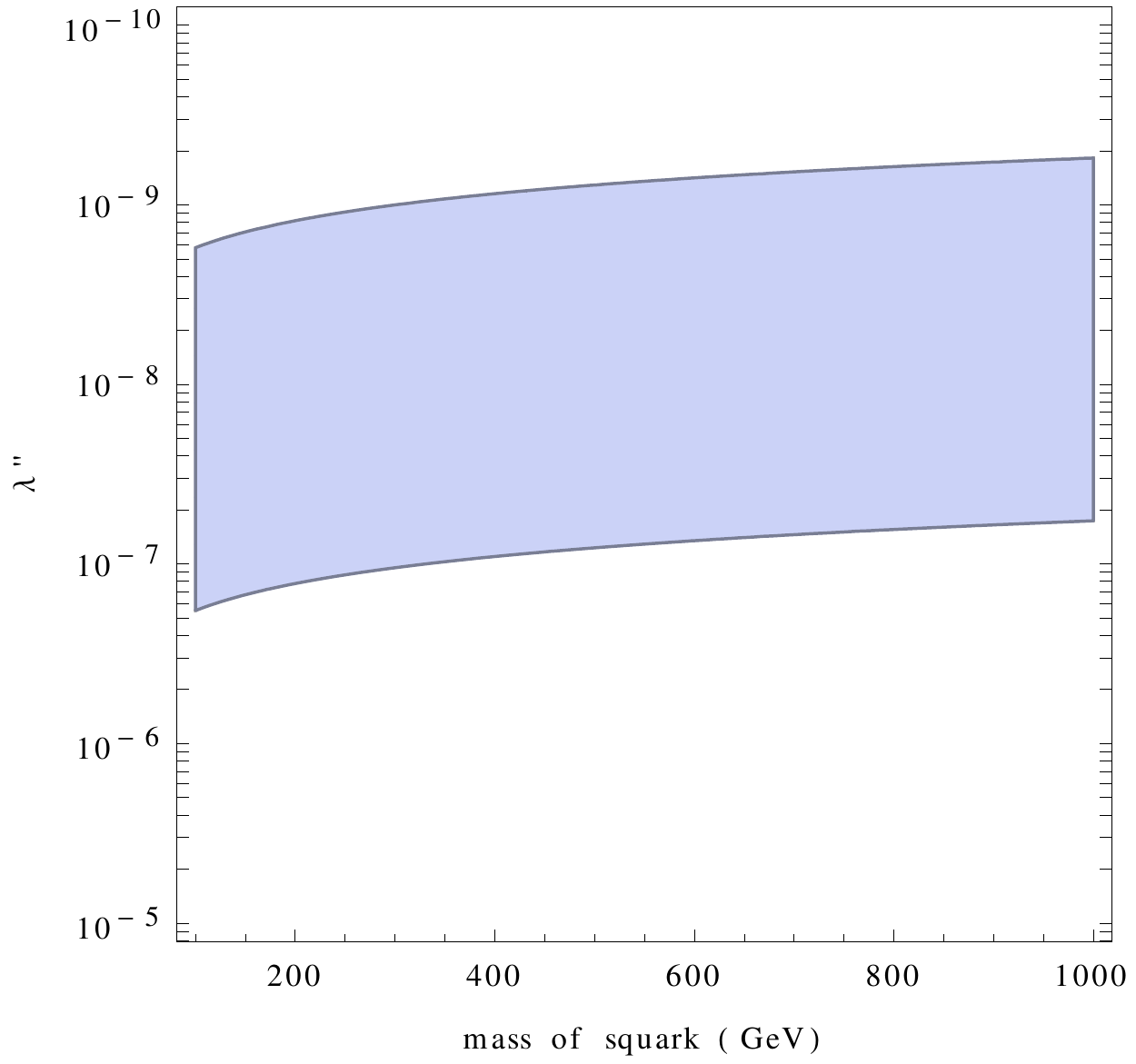} 
\label{HLSP}
}
\subfloat[Assuming hierarchical couplings.]{
\includegraphics[width=0.45\textwidth]{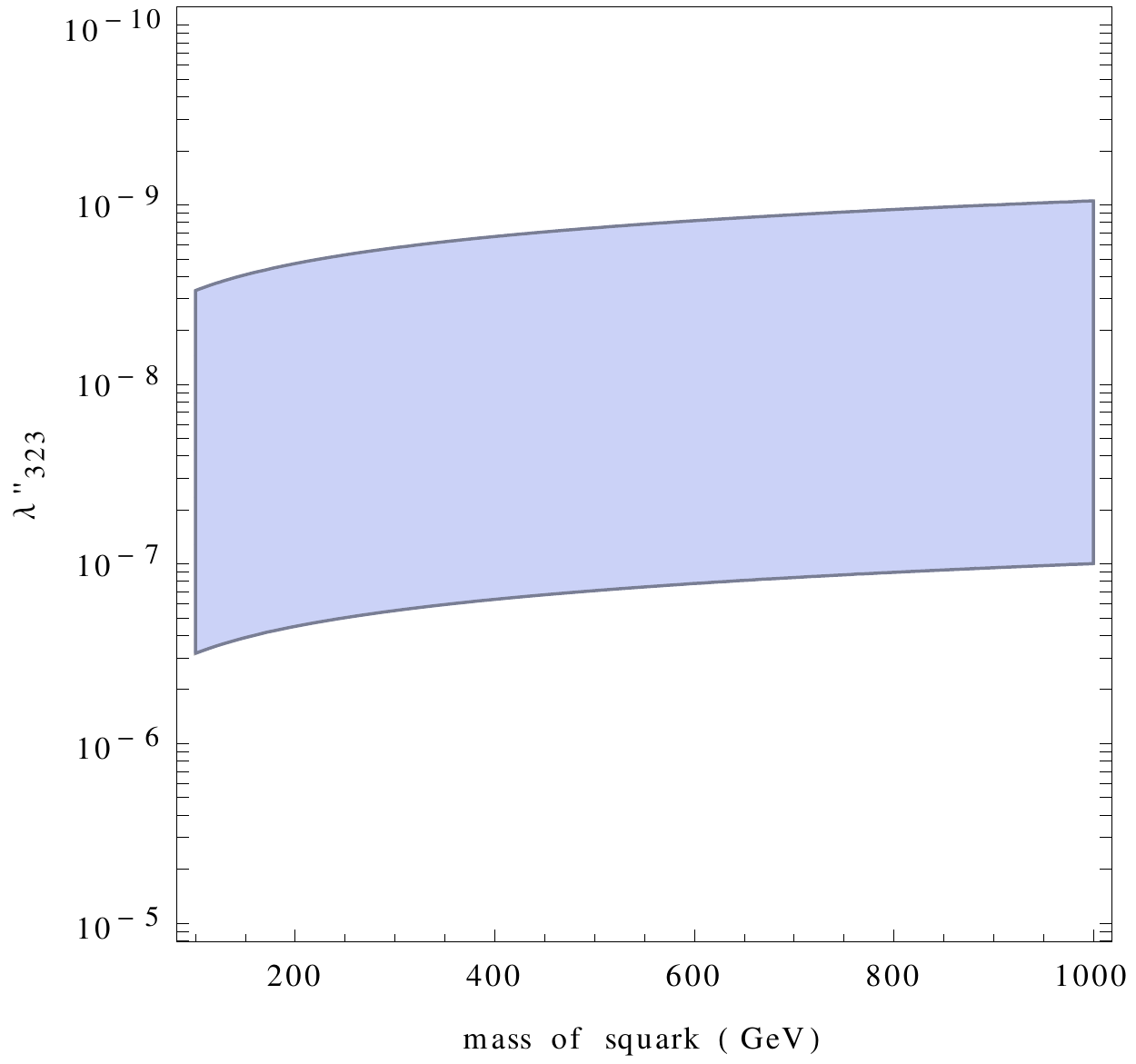}
\label{ULSP}
} 
\caption{Region plot for vertex displacement, $\tau c$, that can be observed in a detector with $\ell_{\text{min}}=1mm$ and $\ell_{\text{max}}=11m$, for a stop LSP.} \label{aa1}
\end{figure}

For a very light stop, $ m_{\tilde{t}} = 100$ GeV, to have an observable vertex displacement in a detector with $l_{min} = 1mm$ and $l_{max}= 11m$, the RPV coupling should be in the range $ 5 \times 10 ^{-9} \leq \lambda '' \leq 6 \times 10^{-7}$ for universal coupling, and for a hierarchical coupling $ 7.5 \times 10^{-9} \leq \lambda _{323}''\leq 8 \times 10^{-7}$ as shown in Figure \ref{aa1}. No channel is suppressed when the RPV couplings have universality, because there is an equal probabilities for the stop to decay via any of the couplings $\lambda _{tbd,tbs,tds}''$ so the branching ratio is equally small for all them. Decays of stops with large masses are more suppressed as shown in Figure \ref{aa1}.

To compare with experimental results that are not explicitly interpreted in this model we calculate the number of events expected in each analysis using equation (\ref{Nevents}) assuming the proper lifetime above. $\gamma \beta \simeq 1$ for most of the stops and prospino cross-section for stops pair production.

To compare with experimental results we calculate the number of events using equation (\ref{Nevents}) with the life time of the LSP stop and the cross section production of the stop from a proton-proton interaction and $\gamma\beta$ is taken to be a constant of order 1.

A pair of stops produced via strong interactions in proton-proton collision would leave the same signature of two isolated, back to back vertexes as in Vd1 and Vd2 in Table\ref{table2}; when the stop decays via a RPV channel. If the stop is long-lived, it may hadronise forming R-hadrons which decay outside the detector, giving the same signature as Od1 in Table\ref{table2}. The stop may decay to a top quark which in turn decays to a muon. In this case the stop will be long-lived and have a muon in the final state, giving the same signature as Vd3 in Table\ref{table2}. 
In addition, the scenario is contrained by dinucleon decay. 

\begin{figure}[h!]
\centering
\subfloat[ Assuming universal couplings.]{
\includegraphics[width=0.55\textwidth]{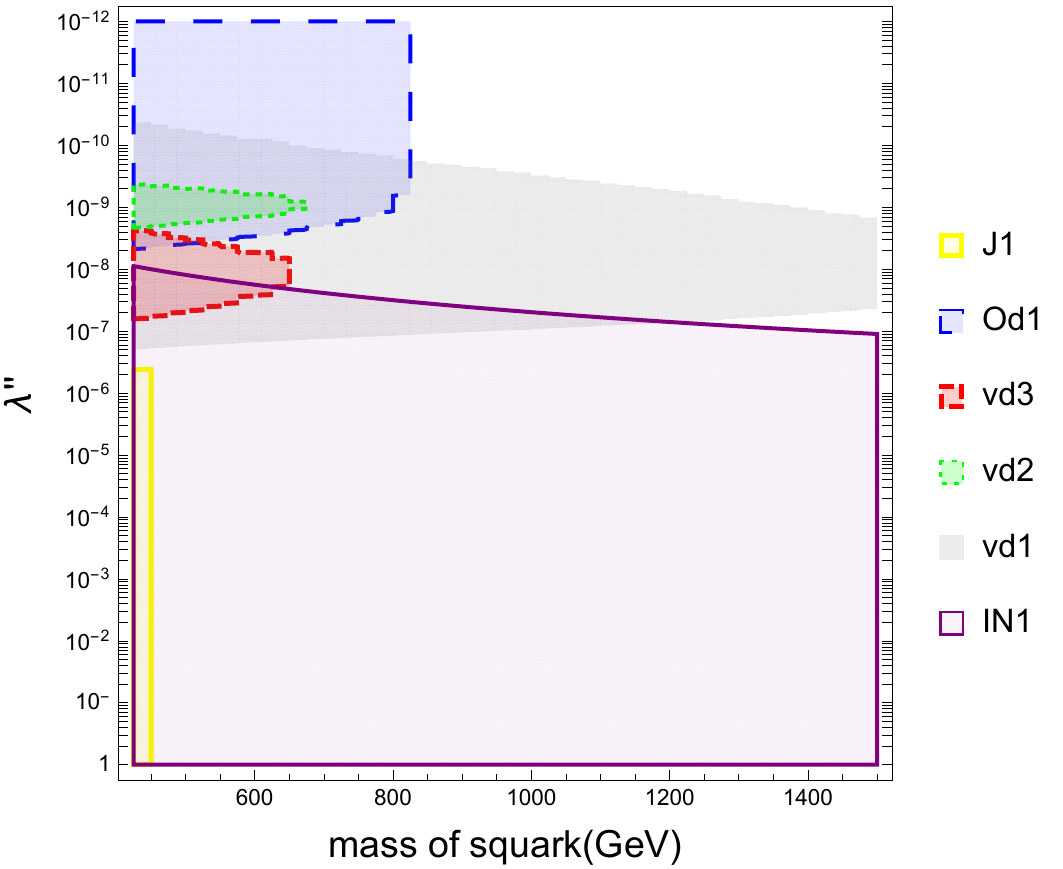}
\label{HLSP}
}
\subfloat[Assuming hierarchical couplings.]{
\includegraphics[width=0.55\textwidth]{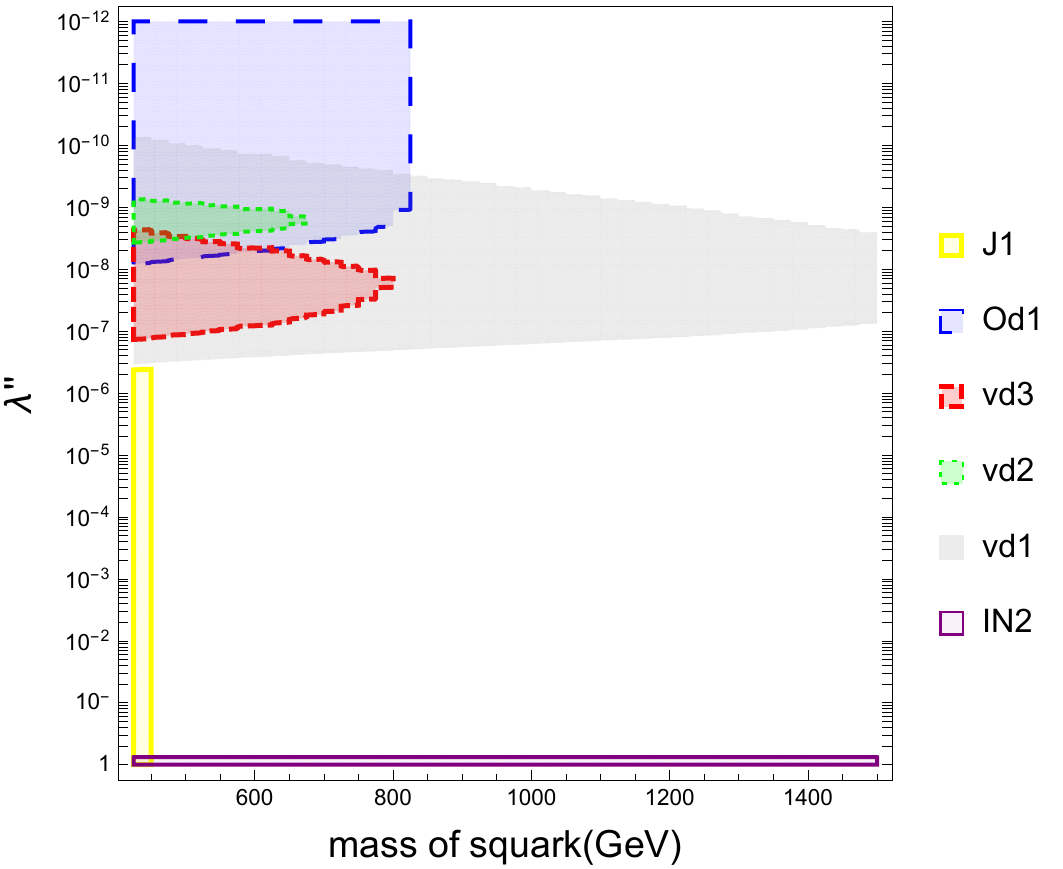}
\label{ULSP}
} 
\caption{Regions excluded by searches in Table\ref{table2} for squark LSP} \label{RPVSLSP}
\end{figure}

These searched together with the bound from dinucleon decay exclude the regions shown in Figure \ref{RPVSLSP}, which shows that for RPV models with stop LSP, $ m_{\tilde{t}} <$ $835$ GeV is robustly excluded irrespective of the RPV coupling strength.  The prompt decay searches cover short lived LSP and long lived LSP are covered by R-hadron searches. These search live a gap, which is closed by the vertex displacement search, which is much more sensetive to large stop masses. 

\subsection{Gluino LSP}

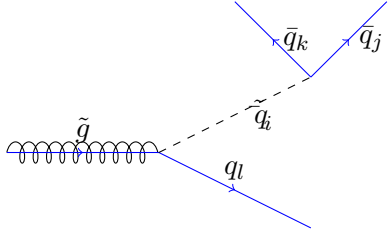
\begin{figure}[h!]
\centering
\begin{tikzpicture}
  \coordinate (a) at (2,2) {};
  \coordinate (b) at (4,2) {};
  \coordinate (c) at (6,1) {};
  \coordinate (d) at (6,3) {}; 
  \coordinate (e) at (7,4) {};
  \coordinate (f) at (5,4) {};
  \draw [gluon] (a) -- (b)node[midway,above] {$\tilde{g}$};
  \draw [particle] (a) -- (b);
  \draw [particle] (b) -- (c)node[midway,above] {$q_l$};
  \draw [antiparticle] (b) -- (d)node[midway,right] {$\tilde{\bar{q}_i}$};
  \draw [particle] (d) -- (e)node[midway,right] {$\bar{q}_j$};
  \draw [particle] (d) -- (f)node[midway,right] {$\bar{q}_k$};
\end{tikzpicture}
\caption[barrier option valuation]{Gluino LSP}
\label{vglsp}
\end{figure}

If the LSP is a gluino, it can decay into three quarks with a virtual squark as shown in Figure \ref{vglsp}. This decay is characterized by up to six jets, absence of missing energy and no isolated leptons.
Due to the small RPV Yukawa couplings the gluino will be expected to have a long lifetime. Assuming that the quark masses are negligible $m_{\tilde{g}},m_{\tilde{q}} \ll m_q$ and $m_{\tilde{q}}\gg m_{\tilde{g}}$, the lifetime of the gluino is
\begin{equation}
\tau _{\tilde{g}} =\frac{384 \pi ^2}{\alpha _s \lambda ^{''2}_{ijk}} \frac{ m^4_{\tilde{q}}}{ m^5_{\tilde{g}}}.
\end{equation}

We will consider four cases:
\begin{itemize}
\item[$\bullet$] \textbf{Case 1:} Assume universality for the squark masses, that is, the gluino has equal probability to decay to the three jets via any virtual squark and that the RPV couplings are also universal, that is, for all $ijk$, $\lambda ''_{ijk}=\lambda ''$. The strongest indirect bound for this case is from the dinucleon decay, $\lambda '' < 10^{-6}$.
\[\tau _{\tilde{g}} =\frac{384 \pi ^2}{18\alpha _s \lambda ^{''2}_{ijk}} \frac{ m^4_{\tilde{q}}}{ m^5_{\tilde{g}}}.\]
\item[$\bullet$] \textbf{Case 2:} Assume that the masses of the squarks are not universal, the stop is lighter than all the other quarks. Also assume that the RPV couplings are universal, that is, $\lambda ''_{3jk}=\lambda ''$. The strongest indirect bound for this case is also from the dinucleon decay.
\[\tau _{\tilde{g}} =\frac{384 \pi ^2}{3\alpha _s \lambda ^{''2}_{3jk}} \frac{ m^4_{\tilde{q}}}{ m^5_{\tilde{g}}}.\]
\item[$\bullet$] \textbf{Case 3:} Assume universality for the squark masses and that the RPV couplings are not universal but have a hierarchy. If we consider the hierarchical RPV couplings for a Minimal Flavor Violating (MFV) model discussed in chapter 2, the largest coupling in this case would come from $\lambda ''_{323}$, we shall consider this coupling for this case. The only relevant indirect bound is perturbativity  $\lambda ''_{323} < 1.2 $ \cite{myrpv}, and
\[\tau _{\tilde{g}} =\frac{384 \pi ^2}{3\alpha _s \lambda ^{''2}_{ijk}} \frac{ m^4_{\tilde{q}}}{ m^5_{\tilde{g}}}.\]
\item[$\bullet$] \textbf{Case 4:} Assume that the stop is lighter than all the other quarks and assume that the RPV couplings have a hierarchy. Considering MFV model the largest coupling is $\lambda ''_{323}$, will consider this coupling for this case. The only relevant indirect bound is perturbativity  $\lambda ''_{323} < 1.2 $ \cite{myrpv}, and
\[\tau _{\tilde{g}} =\frac{384 \pi ^2}{\alpha _s \lambda ^{''2}_{ijk}} \frac{ m^4_{\tilde{q}}}{ m^5_{\tilde{g}}}.\]
\end{itemize}

\begin{figure}[h!]
\centering
\subfloat[ Case 1: Universal Couplings ]{
\includegraphics[width=0.38\textwidth]{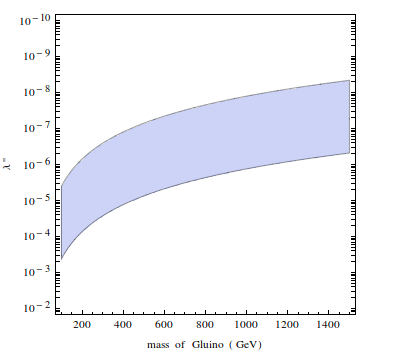}
\label{gdl}
}
\subfloat[ Case4: Hierarchical RPV coupling ]{
\includegraphics[width=0.45\textwidth]{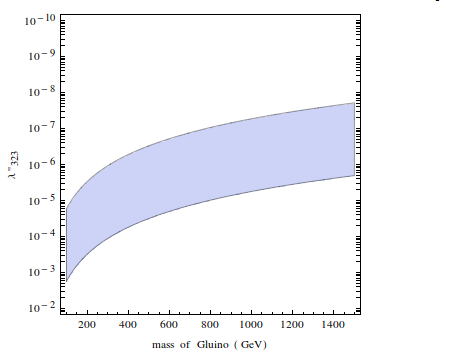}
\label{g4l}
} 
\caption[barrier option valuation]{Region plot for vertex displacement, $\tau c$, that can be observed in a detector with $\ell_{\text{min}}=1mm$ and $\ell_{\text{max}}=11m$, for a gluino LSP decaying via a virtual sqaurk of mass $m_{\tilde{q}}=1.5$ TeV.}
\label{gLSPdl}
\end{figure}

For a gluino LSP to have an observable vertex displacement in a detector with $ 11m \geqslant \ell \geqslant 1mm$ for small gluino masses, $m_{\tilde{g}}=100$ GeV, decaying via a virtual squark of mass $m_{\tilde{q}}=1 TeV$ the allowed couplings are $6 \times 10^{-6} \leq \lambda '' \leq 4 \times 10^{-4}$ if the RPV couplings are universal. Large gluino masses require smaller RPV couplings in order to have observable displacement as shown in Figure \ref{gdl}.

\begin{figure}[h!]
\centering
\subfloat[{\tiny Case 2: Stop lighter than the other squarks and universal RPV coupling. }]{
\includegraphics[width=0.5\textwidth]{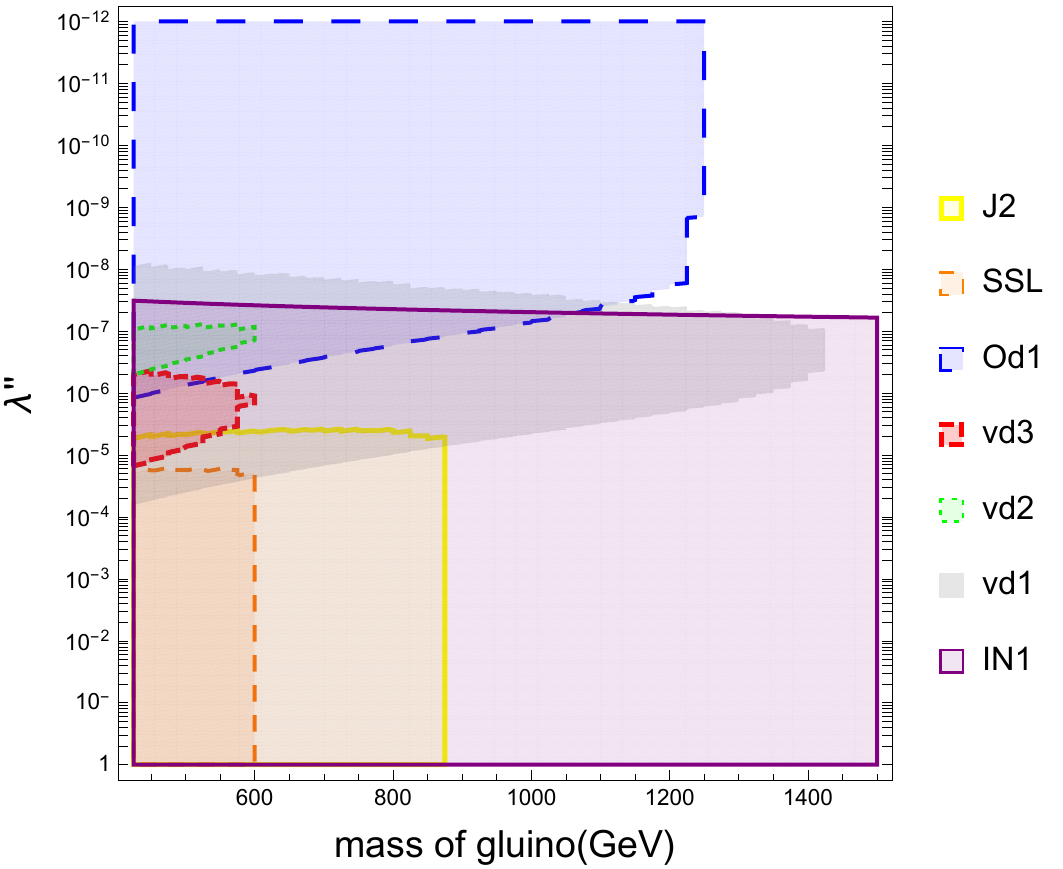}
\label{gelps2}
} 
\subfloat[{\tiny Case 4:  Stop lighter than the other squarks and hierarchical couplings.}]{
\includegraphics[width=0.5\textwidth]{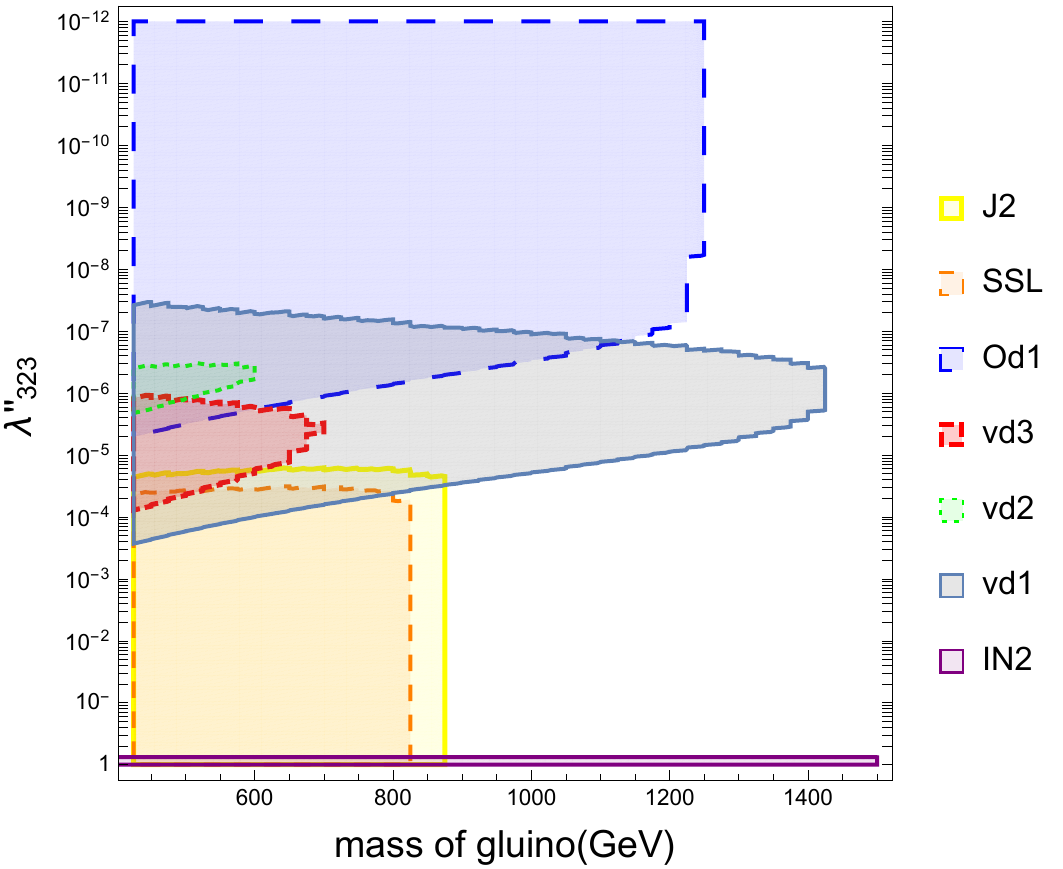}
\label{gelps4}
} 
\caption[barrier option valuation]{Region excluded by the searches in Table\ref{table2} for gluino LSP with virtual squark $m_{\tilde{q}} = 2$ TeV. }
\label{gelsp}
\end{figure}

To compare with experimental results that are not explicitly interpreted in this model we calculate the number of events expected in each analysis using equation (\ref{Nevents}) assuming the proper lifetimes above. $\gamma \beta \simeq 1$ for most of the gluinos and prospino cross-section for gluino pair production. 

A pair of gluinos would leave the same signature of two isolated, back to back vertexes; when the gluino decays via an RPV channel as in Vd1 and Vd2 in Table\ref{table2}. The gluino may decay to a top quark which in turn decays to a muon. In this case the gluino will be long-lived and have a muon in the final state, giving the same signature as Vd3 in Table\ref{table2}. For a pair of gluinos decaying via RPV channel we expect to have the same signature of six jets as in J2 in Table\ref{table2}. If both gluinos decays to top quarks, then the top quarks may both decay to same sign leptons, giving same signature as SSL in Table\ref{table2}. If the gluino is long-lived, it may hadronise forming R-hadrons giving a signature similar to Od1 in Table\ref{table2}.In addition the scenario is constrained by dinucleon decay. \\
 
Figure \ref{gelsp} shows that for RPV model with universal couplings, $ m_{\tilde{g}} <$ $1200$ GeV has been ruled out for all RPV couplings. For and hierarchical couplings, $ m_{\tilde{g}} <$ $975$ GeV have been ruled out for all RPV couplings.

\subsection{Bino LSP}

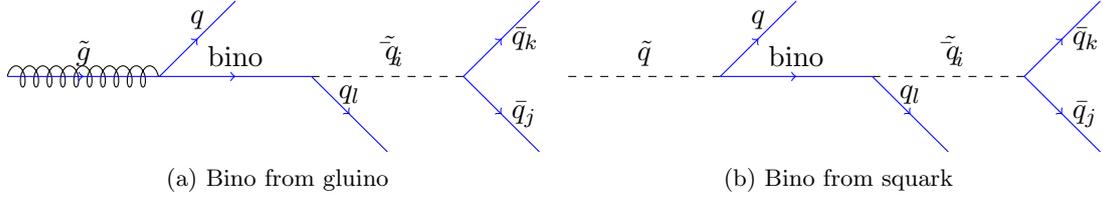
\begin{figure}[h!]
\centering
\subfloat[Bino from gluino]{
\begin{tikzpicture}
  \coordinate (a) at (2,2) {};
  \coordinate (b) at (4,2) {};
  \coordinate (c) at (6,2) {};
  \coordinate (d) at (5,3) {}; 
  \coordinate (e) at (7,1) {};
  \coordinate (f) at (8,2) {};
  \coordinate (h) at (9,1) {};
  \coordinate (g) at (9,3) {};
  \draw [gluon] (a) -- (b)node[midway,above] {$\tilde{g}$};
  \draw [particle] (a) -- (b);
  \draw [particle] (b) -- (d)node[midway,above] {$q$};
  \draw [particle] (b) -- (c)node[midway,above] {bino};
  \draw [antiparticle] (c) -- (f)node[midway,above] {$\tilde{\bar{q}_i}$};
  \draw [particle] (c) -- (e)node[midway,above] {$q_l$};
  \draw [particle] (f) -- (h)node[midway,right] {$\bar{q}_j$};
  \draw [particle] (f) -- (g)node[midway,right] {$\bar{q}_k$};
\end{tikzpicture}
\label{3bino}
}
\subfloat[Bino from squark ]{
\begin{tikzpicture}
  \coordinate (a) at (2,2) {};
  \coordinate (b) at (4,2) {};
  \coordinate (c) at (6,2) {};
  \coordinate (d) at (5,3) {}; 
  \coordinate (e) at (7,1) {};
  \coordinate (f) at (8,2) {};
  \coordinate (h) at (9,1) {};
  \coordinate (g) at (9,3) {};
  \draw [antiparticle] (a) -- (b)node[midway,above] {$\tilde{q}$};
  \draw [particle] (b) -- (d)node[midway,above] {$q$};
  \draw [particle] (b) -- (c)node[midway,above] {bino};
  \draw [antiparticle] (c) -- (f)node[midway,above] {$\tilde{\bar{q}_i}$};
  \draw [particle] (c) -- (e)node[midway,above] {$q_l$};
  \draw [particle] (f) -- (h)node[midway,right] {$\bar{q}_j$};
  \draw [particle] (f) -- (g)node[midway,right] {$\bar{q}_k$};
\end{tikzpicture}
\label{4bino}
} 
\caption[barrier option valuation]{Bino LSP}
\label{bino}
\end{figure}


The LSP can also be a nuetralino. We will consider a specific case when it is a bino, and is produced dominantly from either gluino or squark decay.  Again we assume the gluino (or squark) is produced through a R-parity conserving interaction, so that we have two gluinos (or squarks) that both decay via R-parity conserving decays into a bino LSP, which then decays through a RPV channel into 3 quarks. The lifetime of the bino is,

\begin{align}
\tau &= 9 \frac{ 384 \pi ^2 \cos ^2\theta _W}{ U^{*2}_{il} \alpha \lambda ^{''2}_{ijk}}\frac{m_{\tilde{q}}^4}{m_{\tilde{B}}^5}.
\end{align}

Again we consider four cases as we did for the gluino LSP and the same limits apply as for that case.
\begin{itemize}
\item \textbf{Case 1:} Assume both the mass and the RPV coupling to be universal, summing over all the diagrams the lifetime is
\[ \tau = \frac{1}{5} \frac{ 384 \pi ^2 \cos ^2\theta _W}{ \alpha \lambda ^{''2}}\frac{m_{\tilde{q}}^4}{m_{\tilde{B}}^5}.\]
\item \textbf{Case 2:} Assume the stop is lighter than all the other squarks and the RPV coupling to be universal, the lifetime is
\[ \tau = 3 \frac{ 384 \pi ^2 \cos ^2\theta _W}{ \alpha \lambda ^{''2}_{3jk}}\frac{m_{\tilde{q}}^4}{m_{\tilde{B}}^5}.\]
\item \textbf{Case 3:} Assume the squarks masses are universal and the RPV coupling to be hierarchical, then the lifetime is
\[ \tau = \frac{ 384 \pi ^2 \cos ^2\theta _W}{ \alpha \lambda ^{''2}_{323}}\frac{m_{\tilde{q}}^4}{m_{\tilde{B}}^5}.\]
\item \textbf{Case 4:} Assume both the squarks masses and the RPV coupling to be hierarchical, then the lifetime is
\[ \tau = 9 \frac{ 384 \pi ^2 \cos ^2\theta _W}{ \alpha \lambda ^{''2}_{323}}\frac{m_{\tilde{q}}^4}{m_{\tilde{B}}^5}.\]
\end{itemize}

\begin{figure}[h!]
\centering
\subfloat[{\tiny Case 1: Universal RPV coupling.}]{
\includegraphics[width=0.4\textwidth]{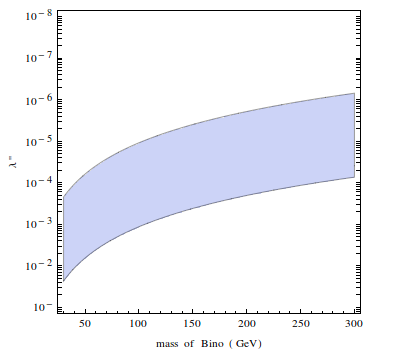}
\label{bgelps}
}
\subfloat[{\tiny Case 4: Hierarchical RPV coupling }]{
\includegraphics[width=0.4\textwidth]{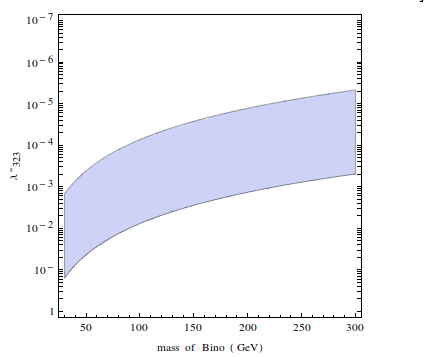}
\label{bselps}
} 
\caption[barrier option valuation]{Region plot for vertex displacement, $\tau c$, that can be observed in a detector with $\ell_{\text{min}}=1mm$ and $\ell_{\text{max}}=11m$, for a bino LSP produced from a gluino, $m_{\tilde{g}}=800$ GeV, with the squark propagator $m_{\tilde{q}}=1$ TeV.}
\label{belsp}
\end{figure}

Heavy binos require smaller couplings in order to be observed in a detector as shown in Figure \ref{belsp}. To calculate the number of events we use equation (\ref{Nevents}) with the lifetime of the bino and the cross-section production of the gluino (or squark) with 
\[\gamma\beta=\frac{1}{3}\frac{m_{\tilde{B}}}{m_{\tilde{g}}}.\]

The two bino LSPs are expected to have isolated and back to back vertices hence has the same signature as Vd1 and Vd2 in Table\ref{table2}. The bino LSP may also have a muon in the final state from a top decay giving the same signature of Vd3 in Table\ref{table2}. If the bino decays outside the detector it leavs a signature of missing energy and two jets like Od2 in Table\ref{table2}. For  decays through $\lambda''_{3jk}$ couplings, the produced top quarks can decay to same sign leptons, giving same signature searched for by the SSL search in Table\ref{table2}.


\begin{figure}[h!]
\centering
\subfloat[ {\tiny Case 1: Universal RPV couplings.}]{
\includegraphics[width=0.5\textwidth]{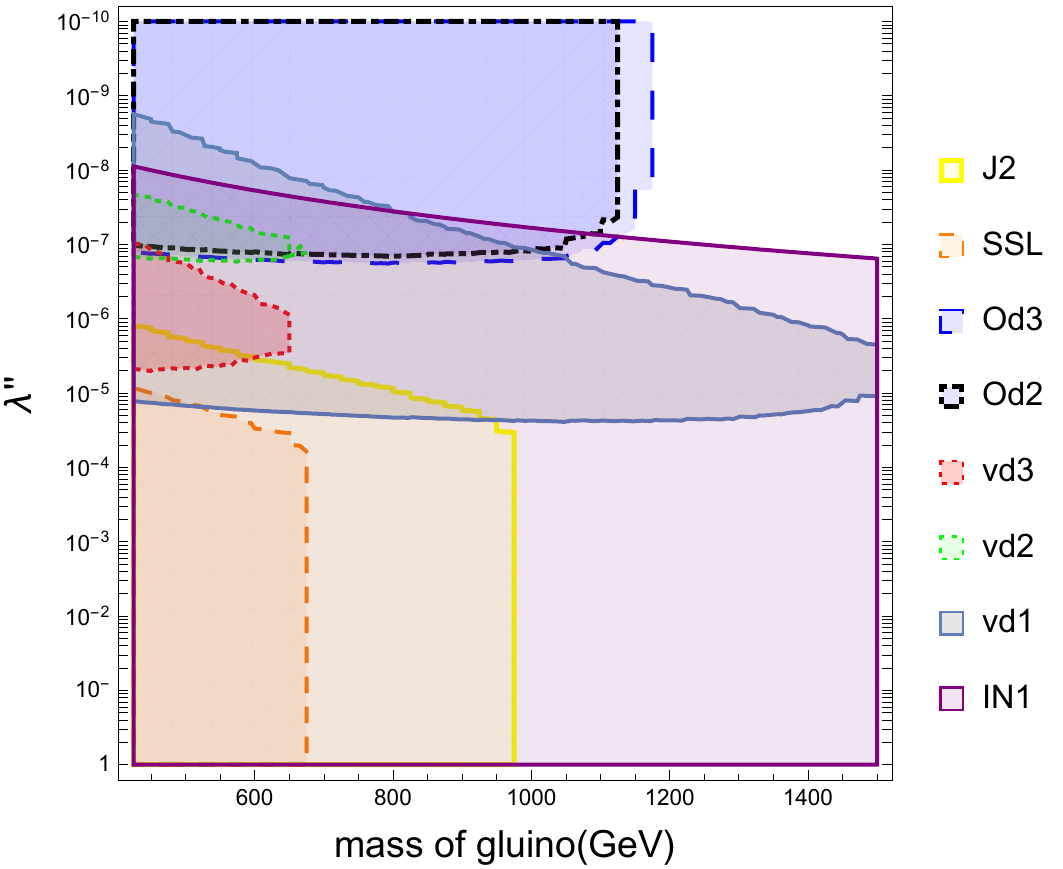}
\label{bgLSP}
} 
\subfloat[{\tiny Case 4: Hierarchical RPV couplings with the stop is lighter than the other quarks.}]{
\includegraphics[width=0.5\textwidth]{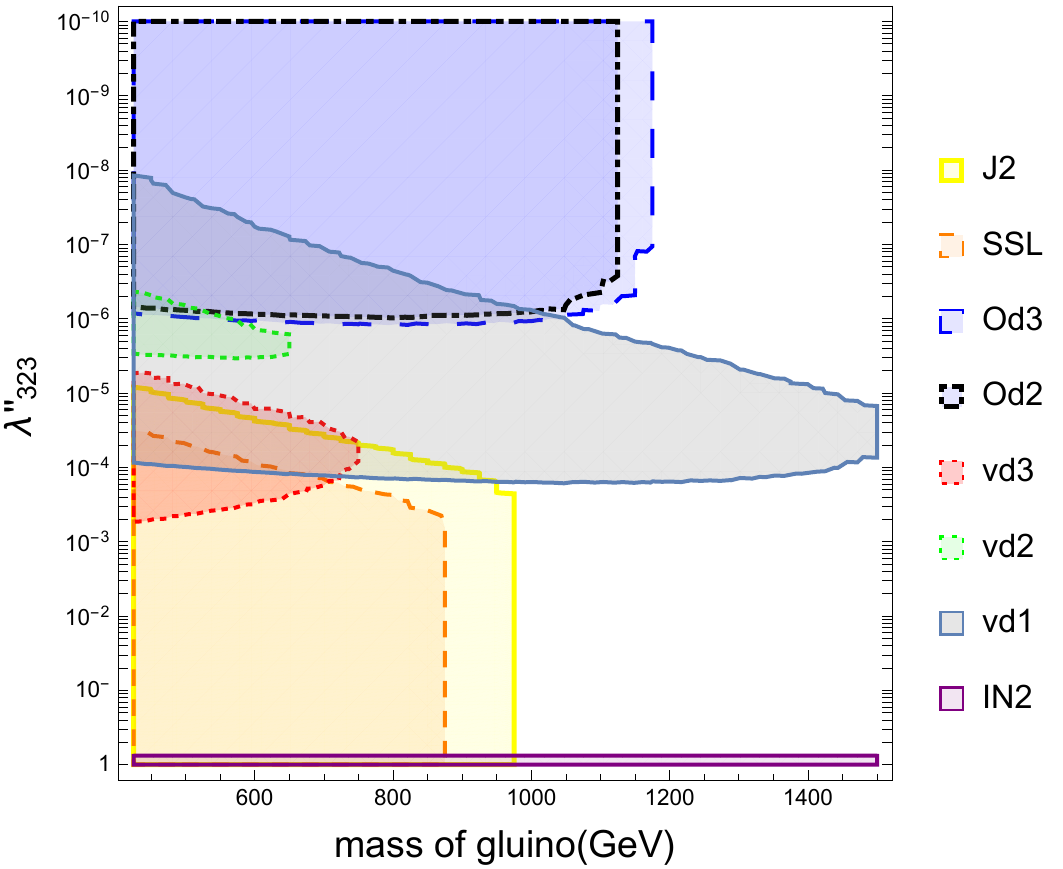}
\label{bbsjLSP}
} 
\caption[barrier option valuation]{Region excluded by the experiments in table\ref{table2} for Bino LSP with $m_{\tilde{B}}=400$ GeV produced from a gluino. The Bino decays via a virtual sqaurk  with $m_{\tilde{q}}=m_{\tilde{g}+100}$ GeV.}
\label{bgLSPv}
\end{figure}

\begin{figure}[h!]
\centering
\subfloat[ {\tiny Case 1: Universal RPV couplings.}]{
\includegraphics[width=0.5\textwidth]{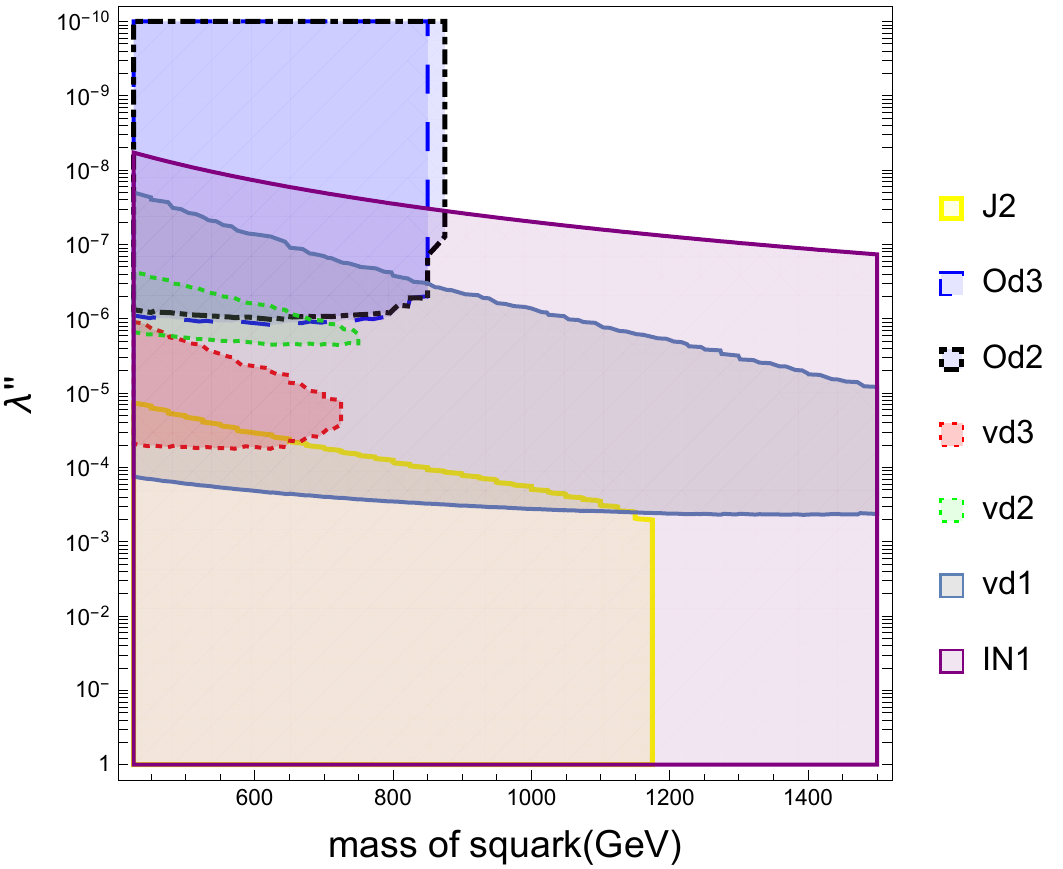}
\label{bgLSP2}
} 
\subfloat[{\tiny Case 4: Hierarchical RPV couplings with the stop is lighter than the other quarks.}]{
\includegraphics[width=0.5\textwidth]{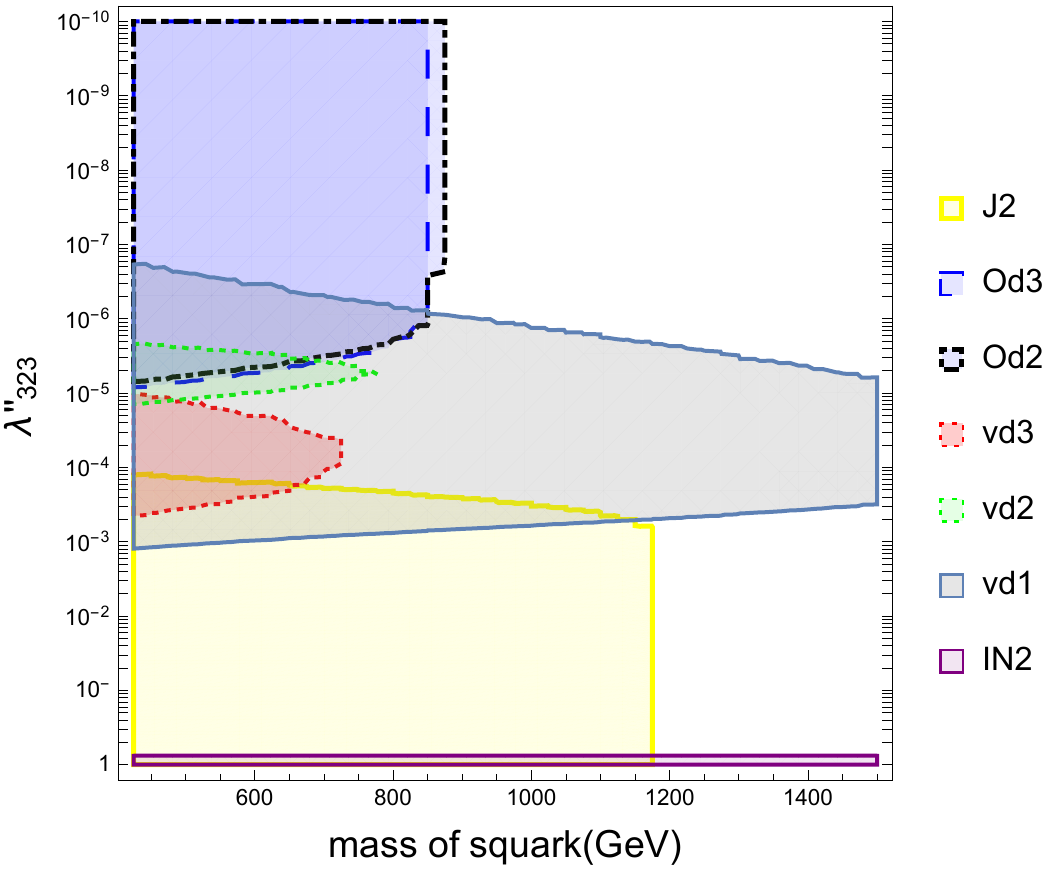}
\label{bbsLSP2}
} 
\caption[barrier option valuation]{Region excluded by the experiments in table\ref{table2} for Bino LSP with $m_{\tilde{B}}=100$GeV produced from a squark. The Bino decays via a virtual sqaurk  with the same mass in and (a) and with $m_{\tilde{q}}=500$ GeV in (b). }
\label{bgLSP2}
\end{figure}

\section{Conclusions}
The search for superpartners in R-parity violating decay signatures complements well the search for superpartners with R-parity conserving decays (or very weak RPV) using large missing energy signals. While searches for \emph{prompt} RPV are quite challenging owing to high backgrounds, the displaced vertex searches have now surpassed the mass sensitivity of missing-energy searches.   

We have considered the sensitivity of a wide variety of LHC searches to three scenarios with baryonic R-parity violation: a third-generation squark LSP, a gluino LSP, and heavy gluinos or squarks decaying into a bino LSP.  In each case, the constraints imply TeV-scale masses for colored superpartners, irrespective of RPV couplings.   

In all cases, displaced vertex searches play an essential role in achieving this sensitivity and are essential to covering the parameter space where small RPV couplings typically give rise to displaced decay vertices.  As Run II of the LHC begins, it will be exciting to see the progress in sensitivity from these three complementary types of searches.


\section{Acknowledgements}
I am very grateful to N. Toro for helpful discussions, suggestions and comments on this work. Research at Perimeter Institute is supported by the Government of Canada through Industry Canada and by the Province of Ontario through the Ministry of Research and Innovation. This work is also part of the research program of the Foundation for Fundamental Research on Matter (FOM), all which is part of the Netherlands Organization for Scientific Research (NWO).

\renewcommand{\bibname}{References}
\nocite{*}
\bibliographystyle{plain}
\bibliography{Rparity}
\addcontentsline{toc}{section}{References}
\end{document}